\documentstyle[aps,twocolumn]{revtex}

\begin{document}
\title{On the self-consistent spin-wave theory of two-dimensional magnets with
impurities}
\author{V.Yu.Irkhin, A.A.Katanin and M.I.Katsnelson$^{*}$}
\address{Institute of Metal Physics, 620219 Ekaterinburg, Russia}
\maketitle

\begin{abstract}
The self-consistent spin-wave theory is applied to investigate the
magnetization distribution around the impurity in isotropic and easy-axis
two-dimensional ferro- and antiferromagents. The temperature dependences of
host magnetization disturbance and impurity magnetization are calculated.
The short-range order in the isotropic case is investigated. Importance of
dynamic and kinematic interactions of spin waves is demonstrated.
\end{abstract}


\section{Introduction}

In connection with extensive investigations of copper-oxide based
superconductors, great attention is paid last time to studying magnetism of
low-dimensional systems. Of particular interest is the problem of
non-magnetic impurities in magnetic hosts. Numerous experimental results
(see, e.g., \cite{La,Zn,Rom}) demonstrate that even small amount of
substitution impurities (Zn, Fe etc.) in CuO$_2$ planes may influence
strongly magnetic properties, e.g. lead to strong suppression of host
magnetization. These facts have stimulated a number of theoretical works
(see, e.g., \cite{Zn,Rom,frustration,swImp}). In particular, the impurity
problem for isotropic two-dimensional (2D) antiferromagnets at $T=0$ was
investigated by the standard spin-wave theory \cite{swImp}. However, the
detailed consideration of the finite temperature situation is absent.
Moreover, the usual spin-wave theory is obviously inapplicable, since this
does not take into account adequately the short-range magnetic order which
is a characteristic feature of low-dimensional magnets.

On the other hand, the impurity problem for three-dimensional (3D) magnets
was investigated within the standard spin-wave theory (see, e.g.\cite{Izumov}%
). It was established that in the case of a weakly coupled magnetic impurity
in a ferromagnet the standard spin-wave approximation is insufficient
already\ at $T\sim T_{imp}$ where $T_{imp}\ll T_C$ \thinspace is the energy
of impurity-host coupling. Inclusion of dynamic and kinematic interaction of
spin waves within the Tyablikov approximation \cite{Izumov} leads in this
case to occurrence of an anomalous temperature dependence of impurity
magnetization. Therefore it is interesting to investigate the impurity
problem for two-dimensional (2D) systems, such as ferro- and
antiferromagnets (FM and AFM) with small anisotropy or interlayer coupling,
which are required to produce a finite value of the magnetic ordering
temperature $T_M$.

In the present paper we consider weakly anisotropic 2D impurity magnetic
crystals with the use of the self-consistent spin-wave theory (SSWT). This
theory was developed to describe thermodynamics of 2D systems \cite
{Arovas,SSWT}, and also successfully applied to quasi-2D \cite
{Quasi-2D,OurFMM} and weakly anisotropic 2D magnets\cite{OurFMM}. An
important advantage of SSWT in comparison with the usual spin-wave theory is
a qualitatively correct description of the strong short-range order above $%
T_M$. Besides that, introducing slave fermions \cite{Jablonski} into SSWT
allows to take into account kinematic interactions of spin waves and
describe systems with not too low $T_M$ values. In the following sections we
treat the cases of different signs of exchange interactions in the host and
between host and impurity.

\section{Ferromagnetic impurity in ferromagnetic host}

The Heisenberg Hamiltonian of a FM crystal with a quadratic lattice,
containing a ferromagnetically coupled impurity at the site $i=0,$ reads 
\begin{equation}
{\cal H}=-\frac 12\sum_{ij}J_{ij}{\bf S}_i{\bf S}_j+{\cal H}_A-H\sum_iS_i^z
\label{H}
\end{equation}
where 
\[
{\cal H}_A=-D\sum_i(S_i^z)^2-\frac 12\sum_{ij}\eta _{ij}S_i^zS_j^z 
\]
is the Hamiltonian of the easy-axis anisotropy, $H$ is external magnetic
field. In the nearest-neighbor approximation the non-zero exchange integrals
are 
\begin{equation}
J_{i,i+\delta }=\left\{ 
\begin{array}{cc}
J^{\prime }, & \text{ }i=0\text{ or }i+\delta =0 \\ 
J, & \,i,i+\delta \neq 0
\end{array}
\right.  \label{J}
\end{equation}
where $\delta $ denotes nearest neighbors, $J>0,\,J^{\prime }>0.$

Following to Ref. \cite{OurFMM} we use in the FM case for $i\neq 0$ the
representation \cite{Jablonski} 
\begin{eqnarray}
S_i^{+} &=&\sqrt{2S}a_i\,,\;S_i^z=S-a_i^{\dagger }a_i-(2S+1)c_i^{\dagger }c_i
\label{BKJ} \\
S_i^{-} &=&\sqrt{2S}(a_i^{\dagger }-\frac 1{2S}a_i^{\dagger }a_i^{\dagger
}a_i)-\frac{2(2S+1)}{\sqrt{2S}}a_i^{\dagger }c_i^{\dagger }c_i  \nonumber
\end{eqnarray}
where $a_i^{\dagger },a_i\ $are the Bose ideal magnon operators and $%
c_i^{\dagger },c_i\ $are the auxiliary Fermi operators at the site $i$ which
take into account the kinematic interaction of spin waves. For $i=0$ one has
to replace in (\ref{BKJ}) $S\rightarrow S^{\prime }$ with $S^{\prime }$ the
impurity spin. Note that in the paper \cite{SSWT} only the Bose operators
were introduced.

To satisfy the condition $\overline{S}_i(H=0)=0$ in the paramagnetic phase
we introduce the Lagrange multipliers $\mu _i$ at each lattice site, which
corresponds to the constraint of the magnon occupation number at $T>T_C$.
These multipliers play the role of a local ``chemical potential'' for the
boson-fermion systems. Introducing $\mu _i$ permits to correct the drawback
of the standard spin-wave theory which is inapplicable at $T>T_C$ since the
magnetization formally becomes negative. Unlike the approach of Ref.\cite
{SSWT}, we do not assume {\it ad hoc} the condition $\overline{S}=0$ in the
ordered phase where we have $\mu _i=0$. Thus in our approach the magnon
number is not conserved at $T<T_C$ and the Bose condensation \cite{SSWT}
does not take place. However, it may be shown that the results of both
approaches are identical at low temperatures where kinematical interactions
are small.

Further we perform decouplings of the quartic forms which occur after
substituting (\ref{BKJ}) into (\ref{H}). Introducing the averages 
\begin{equation}
\xi _{i,i+\delta }=\overline{S}_{i+\delta }+<a_i^{\dagger }a_{i+\delta }>
\label{ksi}
\end{equation}
we derive the quadratic Hamiltonian of the mean-field approximation

\begin{eqnarray}
{\cal H} &=&\sum_{i\delta }\,\xi _{i,i+\delta }J_{i,i+\delta }\left[
a_i^{\dagger }a_i-a_{i+\delta }^{\dagger }a_i+(2S+1)c_i^{\dagger }c_i\right]
\label{R2} \\
&&\ \ \ \ \ \ \ +\sum_i(H-\mu _i)\left[ a_i^{\dagger }a_i+(2S+1)c_i^{\dagger
}c_i\right] +{\cal H}_A  \nonumber
\end{eqnarray}
This Hamiltonian differs from that of the standard spin-wave theory in two
points. First, the averages $\xi _{i,i+\delta }$ are introduced which take
into account the dynamical interaction of spin waves in the lowest Born
approximation (see below). Second, the Fermi operators enter to account the
kinematic interactions of the spin-waves.

Following to Ref. \cite{Anis}, we treat the influence of the magnetic
anisotropy by neglecting quartic forms in ${\cal H}_A$ to obtain 
\begin{eqnarray*}
{\cal H}_A &=&-H_A\sum_iS_i^z \\
&=&-H_A\sum_i\left[ S-a_i^{\dagger }a_i-(2S+1)c_i^{\dagger }c_i\right]
\end{eqnarray*}
with the anisotropy field $H_A$ 
\begin{equation}
H_A=(2S-1)D+S\sum_\delta \eta _{i,i+\delta }  \label{HAnis}
\end{equation}
Note that effects of the true magnetic field $H$ and the field $H_A$ are
different in the paramagnetic phase since chemical potentials $\mu _i$ are
calculated at $H=0$, not $H_A=0$. Thus the field $H$ yields a finite
magnetization at any temperatures, whereas the field $H_A$ induces a slight
shift of $T_C\,\,$only. Thus the field $H_A$ describes correctly the effect
of the easy-axis anisotropy. In the limit $H_A\ll J$ under consideration
effects of single-site and two-site anisotropy are the same, although
concrete expressions for the field $H_A$ in (\ref{HAnis}) are different.

For an ideal crystal $\xi _{i,i+\delta },\mu _i$ do not depend on $i$ and
the diagonalization of the Hamiltonian (\ref{R2}) is easily performed \cite
{SSWT,OurFMM}. At the same time, for the impurity system this is a
complicated task since the unknown dependence $\xi _{i,i+\delta },\,\mu _i$
which is to be determined self-consistently. However, as follows from the
below calculations, $\xi _{i,i+\delta }\,$ and $\mu _i$ practically coincide
with the corresponding quantities for the host, $\xi _M$ and $\mu ,$ except
for nearest neighbors of impurity. Also in the FM phase $\xi _{i,i+\delta }$
as a function of $i$ varies slower than the magnetization, and $\mu _i=0$.
Therefore we may put in (\ref{R2}) 
\begin{equation}
\xi _{i,i+\delta }=\left\{ 
\begin{array}{cc}
\xi , & \text{ }i=0 \\ 
\xi ^{\prime }, & \text{ }i+\delta =0 \\ 
\xi _M, & \text{otherwise}
\end{array}
\right. ,\,\,\,\,\mu _i-\mu =\left\{ 
\begin{array}{cc}
\delta \mu _0, & \text{ }i=0 \\ 
\delta \mu _1, & \text{ }i+\delta =0 \\ 
0, & \text{otherwise}
\end{array}
\right.  \label{Approx}
\end{equation}
Note that $\xi \neq \xi ^{\prime }$ because of non-Hermiticity of the
representation (\ref{BKJ}). Taking into account (\ref{BKJ}) the spin
correlation function of impurity spin with its nearest neighbors has the
form 
\begin{equation}
K\equiv |\,\langle {\bf S}_0{\bf S}_\delta \rangle \,|=\xi \xi ^{\prime }
\label{Corr}
\end{equation}

Under the approximation \ref{Approx} the Hamiltonian (\ref{R2}) takes the
form 
\begin{equation}
{\cal H}={\cal H}_0+V  \label{H0+V}
\end{equation}
$\,$where 
\begin{eqnarray}
{\cal H}_0 &=&J\xi _M\sum_{i\delta }\left[ a_i^{\dagger }a_i-a_{i+\delta
}^{\dagger }a_i+(2S+1)c_i^{\dagger }c_i\right]  \label{H0} \\
&&+(H_A+H-\mu )\sum_i\left[ a_i^{\dagger }a_i+(2S+1)c_i^{\dagger }c_i\right]
\nonumber
\end{eqnarray}
is the standard SSWT Hamiltonian without impurities\cite{SSWT,OurFMM} and 
\begin{eqnarray}
V &=&\left( J^{\prime }\xi -J\xi _M\right) \sum_\delta \left[ a_0^{\dagger
}a_0-a_\delta ^{\dagger }a_0+(2S^{\prime }+1)c_0^{\dagger }c_0\right]
\label{V} \\
&&+\left( J^{\prime }\xi ^{\prime }-J\xi _M\right) \sum_\delta \left[
a_\delta ^{\dagger }a_\delta -a_0^{\dagger }a_\delta +(2S+1)c_\delta
^{\dagger }c_\delta \right]  \nonumber \\
&&+\delta \mu _0b_0^{\dagger }b_0^{}+\delta \mu _1\sum_\delta a_\delta
^{\dagger }a_\delta ^{}  \nonumber
\end{eqnarray}
To diagonalize ${\cal H}$ we introduce the Green's functions

\begin{eqnarray}
G_{ij}^0(\omega ) &=&\ll a_j|a_i^{\dagger }\gg _\omega ^0=\sum_{{\bf q}}%
\frac 1{\omega -E_{{\bf q}}}e^{i{\bf q(R}_i-{\bf R}_j)}  \label{GFunc} \\
G_{ij}(\omega ) &=&\ll a_j|a_i^{\dagger }\gg _\omega  \nonumber
\end{eqnarray}
where the index $0$ means that statististical averages are calculated with $%
H_0,$%
\[
E_{{\bf q}}=\xi _M(J_0-J_{{\bf q}})+H_A+H-\mu ,\;J_{{\bf q}}=2J(\cos
q_x+\cos q_y) 
\]
In the limit $R$ $\gg 1$ we find by using the saddle point approximation
(see, e.g., \cite{Izumov}) 
\begin{equation}
G_{0R}^0(\omega )\sim \left\{ 
\begin{array}{cc}
\exp (i\sqrt{\omega /J\xi _M}R)/\omega ^{1/4}R^{1/2} & 1\ll (\omega
/J)^{1/2}R\,,\,\,\omega \ll 1 \\ 
-\ln (\omega /J) & (\omega /J)^{1/2}R\ll 1
\end{array}
\right.
\end{equation}
The perturbation $V$ can be written in the matrix form

\begin{equation}
V=\sum_{i,,j=0}^4V_{ij}a_i^{\dagger }a_j+\sum_{i=0}^4R_ic_i^{\dagger }c_i^{}
\end{equation}
where the indices $i,j$ enumerate the impurity site and its four nearest
neighbors. From (\ref{V}) we have

\begin{equation}
V=\left( 
\begin{array}{ccccc}
4\varepsilon & \gamma & \gamma & \gamma & \gamma \\ 
\gamma ^{\prime } & \rho & 0 & 0 & 0 \\ 
\gamma ^{\prime } & 0 & \rho & 0 & 0 \\ 
\gamma ^{\prime } & 0 & 0 & \rho & 0 \\ 
\gamma ^{\prime } & 0 & 0 & 0 & \rho
\end{array}
\right) \;,\;\;R=(2S+1)\left( 
\begin{array}{c}
4\varepsilon \\ 
\rho \\ 
\rho \\ 
\rho \\ 
\rho
\end{array}
\right)  \label{Vmatrix}
\end{equation}
\begin{eqnarray*}
\gamma ^{\prime } &=&J^{\prime }\xi -J\xi _M,\,\,\,\varepsilon =\gamma
^{\prime }+\delta \mu _0/4,\; \\
\;\gamma &=&J^{\prime }\xi ^{\prime }-J\xi _M,\,\,\,\,\rho =\gamma +\delta
\mu _1
\end{eqnarray*}
Then we have the expression for the perturbed Green's function \cite{Izumov}:

\begin{equation}
\widetilde{G}(\omega )=[1-\widetilde{G}^0(\omega )V]^{-1}\widetilde{G}%
^0(\omega )  \label{G1}
\end{equation}
where $\widetilde{G}(\omega ),\;\widetilde{G}^0(\omega )$ are submatrices of
matrices $G_{ij}(\omega ),G_{ij}^0(\omega )$ with $i,j$ = 0...4. Further we
calculate the matrix $G$ from (\ref{G1}) and the averages $<a_i^{\dagger
}a_j>$ from the spectral representation. Then we derive from (\ref{BKJ}), (%
\ref{Approx}) the system of self-consistency equations 
\begin{eqnarray}
\xi &=&\overline{S}_1+\int\limits_{-\infty }^{+\infty }\frac{d\omega }\pi
N(\omega )\text{Im}\widetilde{G}_{10}(\omega ),\,  \label{ksis} \\
\,\xi ^{\prime } &=&\overline{S}_0+\int\limits_{-\infty }^{+\infty }\frac{%
d\omega }\pi N(\omega )\text{Im}\widetilde{G}_{01}(\omega )  \nonumber
\end{eqnarray}
where $N(\omega )=1/(\exp (\omega /T)-1)$ is the Bose distribution function.
The integration region in (\ref{ksis}) is in fact $\alpha \leq \omega \leq
2\xi _MJ_0+\alpha ,$ $\alpha =H_A+H-\mu $. The expressions for the site
magnetizations take the form 
\begin{eqnarray}
\overline{S}_0 &=&S^{\prime }-\int\limits_{-\infty }^{+\infty }\frac{d\omega 
}\pi N(\omega )\text{Im}\widetilde{G}_{00}(\omega )+(2S^{\prime }+1)N(E_0) 
\nonumber \\
\overline{S}_1 &=&S-\int\limits_{-\infty }^{+\infty }\frac{d\omega }\pi
N(\omega )\text{Im}\widetilde{G}_{11}(\omega )+(2S+1)N(E_1)  \label{magns}
\end{eqnarray}
where $E_i=(2S_i+1)J\xi _M+\alpha -\delta \mu _i$ is the fermion energy at
the site $i.$

In the case of the pure system ($V=0$) we have $\widetilde{G}=\widetilde{G}%
^0 $ and the values $\overline{S}_i,\,\xi _{i,i+\delta },\,\mu _i$ are
independent of $i,$ so that the system of equations (\ref{ksis}), (\ref
{magns}) reduces to 
\begin{eqnarray}
\xi &=&\xi ^{\prime }=\overline{S}+\frac 1{J_0}\sum_{{\bf k}}J_{{\bf k}}N(E_{%
{\bf k}})  \label{pure} \\
\overline{S} &=&S-\sum_{{\bf k}}N(E_{{\bf k}})+(2S+1)N(E_f)  \nonumber
\end{eqnarray}
where $E_f=(2S+1)J\xi _M+\alpha $. One can see that $\xi $ depends on
temperature due to dynamic magnon-magnon interactions; at low temperatures
the corrections are proportional to $T^{5/2},$ as well as in the Dyson's
theory \cite{Dyson}.

The results of numerical solution of the equations (\ref{pure}) at different
values of $H_A$ are shown on Fig.1. The results of numerical solution of
Eqs. (\ref{pure}) for different $H_A\ $are shown on Fig.1. While the
magnetization is strongly dependent from the value of $H_A$, the dependence $%
\xi (T)$ is the same to calculation accuracy at an arbitrary $H_A/J\ll 1$.
This dependence coincide with those for the ferromagnet in earliar variants
of SSWT \cite{SSWT} at low temperatures. However, at $T\sim J$ the
short-range order parameter $\xi $ demonstrates a sharp decrease rather than
vanishing. Thus introducing the Fermi operators removes the unphysical
transition with vanishing of short-range order parameter. At finite values
of $H_A$ the value of the Curie temperature is finite. At small $H_A\ll J$
we have (cf. \cite{OurFMM}) 
\begin{equation}
T_C=\left\{ 
\begin{array}{cc}
4\pi JS^2/\ln (T/H_A) & 1\ll \ln (J/H_A)\ll 2\pi S \\ 
4\pi JS^2/\ln (\pi ^2J/H_A) & \ln (J/H_A)\gg 2\pi S
\end{array}
\right.
\end{equation}

Now we turn to the consideration of the impurity system. The results of
numerical calculations of magnetizations $\overline{S}_0,\overline{S}_1$ vs.
temperature according to (\ref{ksis}),(\ref{magns}) in the zero magnetic
field are presented in Figs 2,3. In Fig.2 the results of the standard
spin-wave theory (SW) which correspond, in our notations, to $\xi _M=\xi
=\xi ^{\prime }=S,$ fermion occupation numbers $N(E_i)$ being replaced by
zero, and the spin-wave theory with introducing fermions (SWF) are also
presented for comparison. We see that the impurity magnetization has an
anomalous behavior at temperatures $T\sim J^{\prime }.$ On the inset on
Fig.2 this dependence is shown at a different $H_A/J,$ $J^{\prime }/J.$ The
sharp decrease of impurity-site magnetization at $T\sim J^{\prime }$ can be
easily obtained already in the simple mean-field approximation, however the
detailed description of the this behavior requires a more complicated
methods. The standard spin-wave, as well as the SWF solution do not show
this anomaly, so we can conclude that it is caused by both dynamic and
kinematic interactions of spin waves. The situation is similar to the 3D
case where using the Tyablikov approximation results in a strong
modification of the magnetization behavior in this temperature interval \cite
{Izumov}.

In the ground state the disturbance of magnetization is localized at the
impurity site and equals to $S^{\prime }-S.$ To calculate the magnetization
distribution around impurity at finite temperatures we need the full matrix $%
G.$ It may be shown (see, e.g., \cite{Izumov}) that the latter quantity is
given by

\begin{equation}
G=G^0+\widetilde{G}^{0N}V\frac 1{1-\widetilde{G}^0V}\widetilde{G}^{N0}
\label{Spec}
\end{equation}
where $\widetilde{G}^{0N}$ is the submatrix of $G_{ij}^0$ with $i=$ $0..4$, $%
j=0..N,$ and $\widetilde{G}^{N0}$ is the conjugated matrix. Using (\ref{Spec}%
) we can find the averages needed. The results of numerical calculation of
magnetization disturbance for different values of $J^{\prime }/J,$ $H_A/J$
are presented in Fig.4. One can see that at $R>0$ all results are
practically the same, this takes place also in the limiting case with $%
J^{\prime }=0$ (or in the case of vacancy with $S^{\prime }=0$). One can see
that the change of magnetization around the impurity rapidly decreases with
increasing distance from the impurity site, so that the magnetization
disturbance practically vanishes at the distance of $4$ coordination spheres.

In the 2D isotropic magnets where the long-range order at finite
temperatures is absent we have to treat the short-range order parameters $%
\xi _{ij}$ only. The chemical-potential corrections $\delta \mu _i,\,\,i=0,1$
are defined from the condition 
\begin{eqnarray}
&&\ \ \left[ S-<a_i^{\dagger }a_i>-(2S+1)<c_i^{\dagger }c_i>\right] _{H=0}
\label{hc} \\
\ &=&S-\int\limits_{-\infty }^{+\infty }\frac{d\omega }\pi \left. N(\omega )%
\text{Im}\widetilde{G}_{ii}(\omega )\right| _{H=0}  \nonumber \\
&&+\left. (2S+1)N(E_i)\right| _{H=0} 
\begin{array}{c}
=
\end{array}
0  \nonumber
\end{eqnarray}
In zero magnetic field the solution to eqs (\ref{ksis}), (\ref{magns}), (\ref
{hc})\ is not unique. The absense of the unique solution in the paramagnetic
phase is apparently the shortcoming of our approach. Since at small values
of $H$ the solution is unique, it is natural to take $H$ to be small, but
finite (in numerical calculations we have taken $H=0.005J$). Although the
site magnetizations are changed strongly with changing $H$ (the
susceptibility $\chi =\partial \overline{S}/\partial H$ is divergent near $%
T=0$), it may be checked analytically that the derivative $\partial \xi
/\partial H$ remains finite at $H\rightarrow 0,$ so that the values of $\xi
,\xi ^{\prime }$ weakly depend of $H.$ This may be verified also by
numerical calculations.

The numerical procedure is as follows. To find the short-range order
parameters $\xi ,\xi ^{\prime }$ we solve the system of equations (\ref{ksis}%
). At each iteration for given values of $\xi ,\xi ^{\prime },$ the
corrections to the chemical potential $\delta \mu _0,\delta \mu _1$ and
magnetizations $\overline{S}_0,\overline{S}_1$ are determined from eqs (\ref
{hc}) and (\ref{magns}), respectively. Results of numerical calculation of
the short-range order parameters $\xi $,$\xi ^{\prime }$ and correlation
function $K$ for the case of a weakly coupled impurity for the 2D
ferromagnet are shown in Fig.5. One can see that, owing to sharp decrease of 
$\xi ^{\prime },$ the correlations between the impurity site and its nearest
neighbors decrease with temperature more rapidly than those in an ideal
crystal.

\section{Antiferromagnetic impurity in ferromagnetic host}

Further we consider an AFM impurity in FM host $(J>0,\,J^{\prime }<0$ in \ref
{J}). After passing to the local coordinate system at the impurity site, we
have to use the representation 
\begin{eqnarray}
S_0^{+} &=&\sqrt{2S^{\prime }}b_0^{\dagger }\,,\;S_0^z=-S^{\prime
}+b_0^{\dagger }b_0^{}+(2S^{\prime }+1)d_0^{\dagger }d_0^{} \\
S_0^{-} &=&\sqrt{2S^{\prime }}(b_0^{}-\frac 1{2S^{\prime }}b_0^{\dagger
}b_0^{}b_0^{})-2\frac{2S^{\prime }+1}{2S^{\prime }}d_0^{\dagger }d_0^{}b_0^{}
\nonumber
\end{eqnarray}
where $b_0^{\dagger },b_0^{}$ are the Bose operators, $d_0^{\dagger },d_0$
are the Fermi operators. Then, in the mean-field approximation, the
Hamiltonian (\ref{H}) takes the form 
\begin{eqnarray}
{\cal H} &=&\frac 12J\sum_{i,\,i+\delta \neq 0}\xi _{i,i+\delta }\left[
a_i^{\dagger }a_i-a_{i+\delta }^{\dagger }a_i+(2S+1)b_i^{\dagger }b_i\right]
\nonumber \\
&&+|\,J^{\prime }\,|\sum_\delta \left\{ \xi \left[ a_\delta ^{\dagger
}a_\delta -b_0a_\delta +(2S^{\prime }+1)b_\delta ^{\dagger }b_\delta \right]
\right.  \nonumber \\
&&\ \left. +\xi ^{\prime }\left[ b_0^{\dagger }b_0-b_0^{\dagger }a_\delta
^{\dagger }+(2S^{\prime }+1)c_0^{\dagger }c_0\right] \right\}  \label{HF-AF}
\\
&&+\sum_{i\neq 0}(H_A+H-\mu _i)\left[ a_i^{\dagger }a_i+(2S+1)b_i^{\dagger
}b_i\right]  \nonumber \\
&&+(H_A+H-\mu _0)\left[ b_0^{\dagger }b_0+(2S^{\prime }+1)c_0^{\dagger
}c_0\right]  \nonumber
\end{eqnarray}
where 
\begin{eqnarray*}
\xi &=&\overline{S}_0+<d_0^{\dagger }a_\delta ^{\dagger }> \\
\xi ^{\prime } &=&\overline{S}_1+<a_\delta d_0>
\end{eqnarray*}
As in ferromagnetic case, we use the approximation $\xi _{i,i+\delta }\simeq
\xi _M$ ($i,i+\delta \neq 0$). To diagonalize (\ref{HF-AF}) we introduce,
following to \cite{Izumov}, the ``hole'' creation and annihilation operators 
$a_0^{\dagger },$ $a_0$ by the canonical transformation 
\[
a_0=d_0^{\dagger },\,\,\,a_0^{\dagger }=-d_0 
\]
As well as in the case of FM impurity, we use the approximation (\ref{Approx}%
). We introduce also the Green's functions (\ref{GFunc}) and represent the
Hamiltonian as (\ref{H0+V}) with the parameters of the matrix $V$ (\ref
{Vmatrix}) 
\begin{eqnarray*}
\gamma ^{\prime } &=&-J^{\prime }\xi +J\xi _M,\,\,\,\varepsilon =-J^{\prime
}\xi -J\xi _M+\delta \mu _0/4, \\
\;\;\gamma &=&J^{\prime }\xi ^{\prime }+J\xi _M,\,\,\,\,\rho =J^{\prime }\xi
-J\xi _M+\delta \mu _1
\end{eqnarray*}
Then we have the same equation (\ref{G1}) for the full Green's function as
in the case of FM impurity, the self-consistency equations also has the same
form (\ref{ksis}), (\ref{magns}). Unlike the FM impurity case, the full
Green's function has a pole at $\omega =-\omega _0<0$ \cite{Izumov}. To take
into account the contribution from this pole to the averages needed we
deform the integration path in the spectral representation for the Green's
function in the complex plane: 
\begin{eqnarray}
\ &<&a_j^{\dagger }a_i>=\frac 1\pi \int\limits_{-\infty }^{+\infty }d\omega
N(\omega )\text{Im}G_{ij}(\omega )  \label{Spectr} \\
\ &=&\int\limits_C\frac{d\omega }{2\pi i}N(\omega )G_{ij}(\omega )-TG_{ij}(0)
\nonumber
\end{eqnarray}
The contour $C$ is selected in such a way that all singularities of $%
G(\omega )$ lie inside $C,$ but all the frequencies $\omega _n=2\pi
nT\;(n\neq 0)$ lie outside it. The last term in (\ref{Spectr}) corresponds
to the contribution from $\omega =0$ which is to be subtracted explicitly.

Results of numerical solution of Eqs. (\ref{ksis}), (\ref{magns}) with using
of (\ref{Spectr}) for different values of impurity-host coupling and $H=0$
are presented in Figs.6,7.

The AFM impurity induces the disturbance of host magnetization already at $%
T=0.$ Using the sum rule 
\begin{equation}
\pi \sum_i\text{Im}\left[ G_{ii}(\omega +i\delta )-G_{ii}^0(\omega +i\delta
)\right] =\frac \partial {\partial \omega }\text{Im}\ln \det [1-G_0(\omega
)V]  \label{det}
\end{equation}
which follows from (\ref{Spec}) and taking into account that $\det
[1-G_0(\omega )V]$ has a zero at $\omega =-\omega _0$ we obtain 
\begin{equation}
<b_0^{+}b_0>_{T=0}=\sum_{i>0}<a_i^{+}a_i>_{T=0}
\end{equation}
so that the total disturbance of magnetization equals to $S+S^{\prime }$ 
\cite{Izumov}. The distribution of magnetization around the impurity site is
shown in Fig. 8. At large $R$ the contribution from the pole $\omega
=-\omega _0$ gives main contribution to the magnetization disturbance which
is proportional to $\exp \left( -R\sqrt{\omega _0/J}\right) /R$ and differs
from that in the $3D$ case \cite{Izumov} by preexponential factor only.

\section{The case of an antiferromagnetic host}

Now we consider an antiferromagnet with the Hamiltonian

\[
{\cal H}=-\frac 12\sum_{ij}J_{ij}{\bf S}_i{\bf S}_j+{\cal H}_A+H_{{\bf Q}%
}\sum_ie^{i{\bf QR}_i}S_i^z 
\]
with $J_{ij}<0,\,\eta _{ij}<0,\,{\bf Q=}(\pi ,\pi ),$ $H_{{\bf Q}}$ is the
staggered magnetic field. In the case of two sublattices $A,B$ and for the
antiferromagnetically coupled impurity spin in the $A$ sublattice we have to
use the representation

\begin{eqnarray}
S_i^{+} &=&\sqrt{2S}a_i^{}\,,\;S_i^z=S-a_i^{\dagger
}a_i^{}-(2S+1)c_i^{\dagger }c_i^{} \\
S_i^{-} &=&\sqrt{2S}(a_i^{\dagger }-\frac 1{2S}a_i^{\dagger }a_i^{\dagger
}a_i^{})-\frac{2(2S+1)}{2S}a_i^{\dagger }c_i^{\dagger }c_i^{}  \nonumber
\end{eqnarray}
for $i\in A$ and 
\begin{eqnarray}
S_i^{+} &=&\sqrt{2S}b_i^{\dagger }\,,\;S_i^z=-S+b_i^{\dagger
}b_i^{}+(2S+1)d_i^{\dagger }d_i^{} \\
S_i^{-} &=&\sqrt{2S}(b_i^{}-\frac 1{2S}b_i^{\dagger }b_i^{}b_i^{})-\frac{%
2(2S+1)}{2S}d_i^{\dagger }d_i^{}b_i^{}  \nonumber
\end{eqnarray}
for $i\in B$ where $a_i^{\dagger },a_i^{}$ and $b_i^{\dagger },b_i^{}$ are
the Bose operators, $c_i^{\dagger },c_i\,$ and $d_i^{\dagger },d_i$ are the
Fermi operators. After standard decoupling the Hamiltonian takes the form 
\begin{eqnarray}
{\cal H} &=&\sum_{i\in A,\delta }|J_{i,i+\delta }|\,\xi _{i,i+\delta }\left[
a_i^{\dagger }a_i^{}-b_{i+\delta }^{}a_i^{}+(2S+1)c_i^{\dagger }c_i^{}\right]
\label{HAF} \\
&&+\sum_{i\in B,\delta }\,|J_{i,i+\delta }|\widetilde{\xi }_{i,i+\delta
}\left[ b_{i+\delta }^{\dagger }b_{i+\delta }^{}-b_{i+\delta }^{\dagger
}a_i^{\dagger }+(2S+1)c_i^{\dagger }c_i^{}\right]  \nonumber \\
&&+\sum_{i\in A}(H_A+H-\mu _i)\left[ a_i^{\dagger }a_i^{}+(2S+1)c_i^{\dagger
}c\right]  \nonumber \\
&&+\sum_{i\in B}(H_A+H-\mu _i)\left[ b_i^{\dagger }b_i^{}+(2S+1)d_i^{\dagger
}d_i\right]  \nonumber
\end{eqnarray}
where 
\begin{equation}
H_A=(2S-1)D+S\sum_i|\eta _{i,i+\delta }|
\end{equation}
and 
\begin{equation}
\xi _{i,i+\delta }=\overline{S}_{i+\delta }+<b_{i+\delta }^{\dagger
}a_i^{\dagger }>,\,\widetilde{\xi }_{i,i+\delta }=\overline{S}%
_i+<a_i^{}b_{i+\delta }^{}>  \label{ksia}
\end{equation}
For the correlation function $K$ we have the same expression (\ref{Corr}) as
in the FM case with $\xi =\xi _{01},\xi ^{\prime }=\widetilde{\xi }_{01}$.
To diagonalize (\ref{HAF}) we introduce the operators

\[
A_i=\left\{ 
\begin{array}{cc}
a_i & i\in A \\ 
b_i^{\dagger } & i\in B
\end{array}
\right. 
\]
and the Green's functions: 
\begin{equation}
\widehat{G}_{ij}^0(\omega )=\ll A_i^{}|A_j^{\dagger }\gg _\omega
=G_{ij}^0(\Omega )\times \left\{ 
\begin{array}{cc}
r, & \text{ }i,j\in A\text{ } \\ 
r^{-1}, & \text{ }i,j\in B \\ 
1, & \text{otherwise}
\end{array}
\right.  \label{GGG}
\end{equation}
where 
\begin{eqnarray*}
\,r &=&\left( \frac{\lambda +\omega }{\lambda -\omega }\right)
^{1/2},\,\,\,\,\Omega =\lambda -\sqrt{\lambda ^2-\omega ^2},\, \\
\,\lambda &=&|\,J_0\,|\xi _M+H_A+H-\mu
\end{eqnarray*}

Using the approximation (\ref{Approx}) we get the same expression for the
Green's function (\ref{G1}) with $\widetilde{G}^0(\omega )$ being the $%
5\times 5$ submatrix of $\widehat{G}_{ij}^0(\omega ),$ and analogously for $%
\widetilde{G}(\omega )$. In this designations the self-consistent equations
for the site magnetizations and short-range order parameters has the same
forms as in FM case (\ref{magns}), (\ref{ksis}). In the case of the pure
system we now have $\widehat{G}=\widehat{G}_0$ so that 
\begin{eqnarray*}
\xi &=&\xi ^{\prime }=\overline{S}+\frac 1{2J_0}\sum_{{\bf k}}\frac{\xi J_{%
{\bf k}}^2}{E_{{\bf k}}^{\text{AF}}}\coth \frac{E_{{\bf k}}^{\text{AF}}}{2T}
\label{pureAF} \\
\overline{S} &=&S-\sum_{{\bf k}}\frac{\xi J_{{\bf k}}}{2E_{{\bf k}}^{\text{AF%
}}}\coth \frac{E_{{\bf k}}^{\text{AF}}}{2T}+(2S+1)N(E_f)
\end{eqnarray*}
where 
\[
E_{{\bf k}}^{\text{AF}}=\sqrt{(|J_0|\xi _M+\alpha )^2-(J_{{\bf k}}\xi _M)^2} 
\]

The results of numerical calculations for the AFM impurity system case are
shown and compared with those for the FM case in Fig.9. In the case of an
impurity spin, which is weakly coupled to the host, the behavior of
magnetization in AFM and FM situations is very close, except for the region
near the magnetic ordering temperature ($T_N>T_C$ because of quantum
fluctuations). At the same time, the nearest-neighbor magnetizations are
strongly different and demonstrate a behavior, typical for the corresponding
hosts. One can also see that SSWT leads to unambiguous result at low $T$,
where the impurity magnetization turns out to be greater than the host one.
The difference between magnetizations of impurity and host increases with
lowering value of $J^{\prime }/J$ and decreases with increasing temperature.
Thus SSWT predicts strong influence of quantum fluctuations on magnetization
in the case of weakly magnetic impurities.

The results of calculating the short-range order parameters $\xi ,\xi
^{\prime }$ and the correlation function $K$ in the isotropic case $H_A=0$
are presented in Fig. 10. We use the same procedure as in the FM case. The
parameter $\xi $ has a non-monotonic temperature dependence. At the same
time, the temperature dependence of the correlation function of the impurity
spin with its nearest neighbors is monotonic and more rapid than that for
correlation functions between spins in the host.

To calculate the total magnetization disturbance we use the sum rule for the
Green's functions (\ref{GGG}) 
\begin{equation}
\pi \sum_i(-1)^i\text{Im}\left[ \widehat{G}_{ii}(\omega +i\delta )-\widehat{G%
}_{ii}^0(\omega +i\delta )\right] =\frac \partial {\partial \omega }\text{Im}%
\ln \det [1-\widehat{G}^0(\omega )V]
\end{equation}
Since det$(1-\widehat{G}^0V)$ has no zeros at $\omega <0$, we obtain at $T=0$%
\begin{equation}
\delta M=S^{\prime }-S-\left. \frac 1\pi \text{Im }\ln \det (1-\widehat{G}%
^0V)\right| _{-\infty }^0=S^{\prime }-S  \label{SS1}
\end{equation}
This result is valid also for a vacancy if we put $S^{\prime }=0.$ For a
ferromagnetically coupled impurity we have to replace in (\ref{SS1}) $%
S^{\prime }\rightarrow -S^{\prime }.$

The distribution of magnetization around impurity in the ground state of a
2D isotropic antiferromagnet is shown in Fig.11. The magnetization of each
sublattice decreases, so that corrections to the host magnetization have
alternating signs. The values of sublattice magnetization disturbance are
close to those in the spin-wave theory \cite{swImp}. At large $R,$ main
contribution to the disturbance of sublattice magnetization comes from
frequencies $\omega \ll J.$ Expanding (\ref{Spec}), (\ref{GGG}) up to first
order in $\omega /J$ we derive 
\begin{equation}
\delta <A_0^{\dagger }A_i>\sim 1/R_i^3
\end{equation}
\ Note that in the 3D case this quantity demonstrates a more rapid decrease $%
1/R^4,$ which may be obtained in the same manner.

\section{Conclusions}

To conclude, we have investigated 2D magnets with impurities for different
signs of exchange integrals within the framework of self-consistent
spin-wave theory \cite{Arovas,SSWT}. This theory permits to calculate both
magnetization distribution and the correlation functions (short-range order
parameters). For $T=0$ modifications of the results of the standard
spin-wave theory are small. At the same time, for finite temperatures,
corrections owing to dynamic and kinematic interactions of spin waves turn
out to be important. It should be stressed that despite the absence of
long-range order in the isotropic 2D magnets at $T>0$, the temperature
dependence of the imurity-host correlation function $K\,$(\ref{Corr}) is
similar to that in the 3D case, although in the latter case main
contribution to $K$ equals to $\overline{S}_0\overline{S}_\delta $.

The distribution of magnetization in the ground state was investigated in
detail. In the nearest-neighbor approximation considered, the host
magnetization disturbance decreases rapidly with distance from impurity, and
the total change of magnetic moment equals to $-S\pm S^{\prime }$ depending
on the sign of $J^{\prime }.$ More interesting situations occur in the case
of the long-range exchange. So, in the case of FM impurity in the FM host
with sufficiently strong negative next-nearest impurity-host exchange $%
J^{\prime \prime },$ the total magnetization change equals 
\begin{equation}
\delta M=S^{\prime }-S-2z_2S  \label{M1}
\end{equation}
with $z_2$ the corresponding coordination number. In the case of FM impurity
in the AFM host with large positive $J^{\prime \prime }$ we have 
\begin{equation}
\delta M=S^{\prime }-S+2z_2S.  \label{M2}
\end{equation}

It would be of interest also to investigate the problem of a current carrier
in the AFM host within a similar approach (e.g., within the $t-J$ model, cf.%
\cite{Arovas}).

The work was supported in part by Grant No.95-056 from the State Scientific
and Technical Program ``Actual Researches in Condensed Matter Physics",
subprogram ``Superconductivity".

\newpage

{\sc Figure captions.}

Fig.1 The temperature dependences of the magnetization for the pure 2D
ferromagnet with $S=1/2$, $H_A/J=10^{-2}$ (solid line), $H_A/J=10^{-3}$
(dashed line) and short-range order parameter $\xi $ (dot-dashed line which
is the same for all three cases: $H_A/J=0,10^{-3},10^{-2}$).

Fig.2 The temperature dependence of the magnetizations for impurity site $%
\overline{S}_0,$ and for nearest-neighbor sites $\overline{S}_1$ ($%
S=S^{\prime }=1/2,\,H_A/J=10^{-3},\,J^{\prime }/J=0.15$). The results of the
magnetization for the impurity site in the standard (non-self-consistent)
spin-wave approach without (SW) and with (SWF) introducing Fermi operators
(see (\ref{BKJ})) are presented for comparison. On the inset the temperature
dependence of the impurity site magnetization $\overline{S}_0$ at $%
H_A/J=10^{-3}\,,\,J^{\prime }/J=0.15$ (solid line), $H_A/J=10^{-3},\,J^{%
\prime }/J=0.05$ (short-dashed line), $H_A/J=10^{-2},\,J^{\prime }/J=0.15$
(long-dashed line).

Fig.3 The temperature dependence of the short-range order parameters $\xi
,\xi ^{\prime }$ (solid lines, left scale) and correlation function $%
K=\langle {\bf S}_0{\bf S}_\delta \rangle $ (dashed line, right scale) for
the same parameter values as in Fig.2. Arrow shows the value of the Curie
temperature.

Fig.4. The distribution of magnetization around impurity for the same
parameter values as in Fig.2, $T=0.3J$. Arrows show the value of
magnetization disturbance at the impurity site ($R=0$).

Fig.5. The temperature dependence of the parameters $\xi ,\xi ^{\prime }$
and spin correlation function $K=\langle {\bf S}_0{\bf S}_\delta \rangle $
in the isotropic 2D ferromagnet with $S=S^{\prime }=1/2,\,\,J^{\prime
}/J=0.15$. For comparison, the corresponding short-range order parameter in
the ideal crystal, $\xi _M,$ is shown.

Fig.6. The temperature dependence of the magnetizations for impurity site $%
\overline{S}_0,$ and for nearest-neighbor sites $\overline{S}_1$ in the case
of antiferromagnetic impurity in the ferromagnetic host with $S=S^{\prime
}=1/2,\,H_A/J=10^{-3},\,J^{\prime }/J=-0.15.$

Fig.7. The temperature dependence of the magnetizations for impurity site $%
\overline{S}_0,$ and for nearest-neighbor sites $\overline{S}_1$ in the case
of antiferromagnetic impurity in the ferromagnetic host with $S=S^{\prime
}=1/2,\,H_A/J=10^{-3},\,J^{\prime }/J=-1.$

Fig.8. The distribution of magnetization around antiferromagnetic impurity
in the $2D$ isotropic ferromagnet at $T=0,$ $J^{\prime }/J=0.15$ (solid
line), $J^{\prime }/J=0.05$ (dashed line).

Fig.9. The temperature dependence of magnetizations for impurity site $%
\overline{S}_0,$ and for nearest-neighbor sites $\overline{S}_1$, in an
antiferromagnet with $S=S^{\prime }=1/2,\,H_A/J=10^{-3},\,J^{\prime }/J=0.15$%
. Dashed lines show the corresponding results for a ferromagnet (Fig.2).

Fig.10. The temperature dependence of the short-range order parameters $\xi
,\xi ^{\prime }$ and correlation function $K=-\langle {\bf S}_0{\bf S}%
_\delta \rangle $ for the same parameter values as in Fig.9. Arrow shows the
value of the Neel temperature.

Fig.11. The distribution of magnetization around impurity in the $2D$
isotropic antiferromagnet at $T=0$. The inset shows the picture at large $R.$

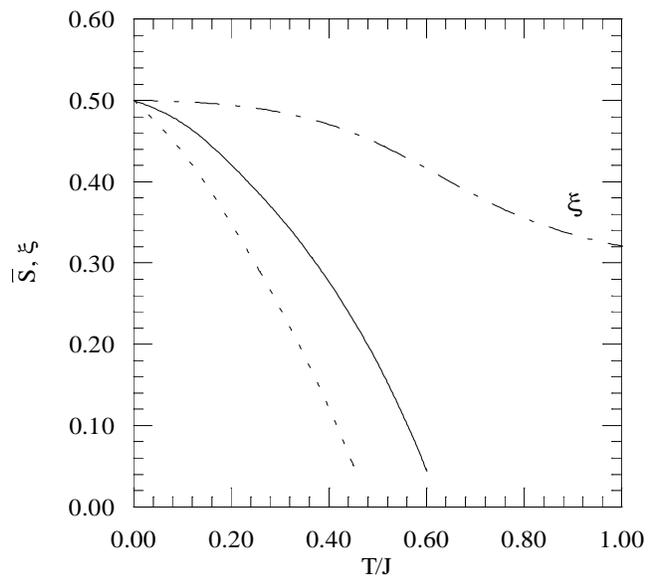
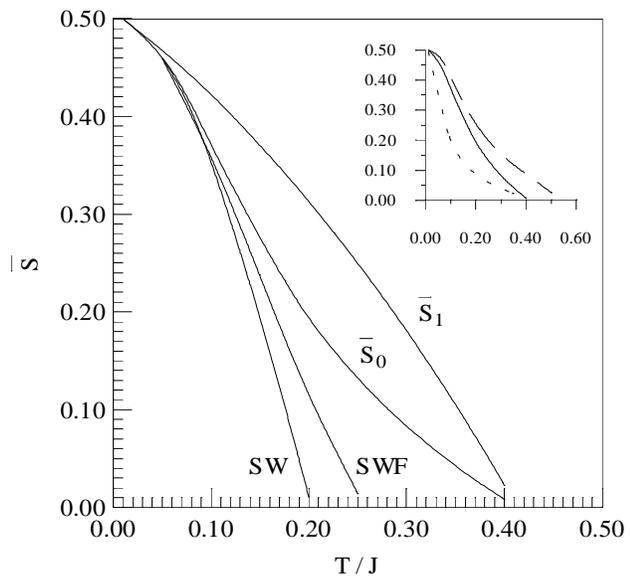
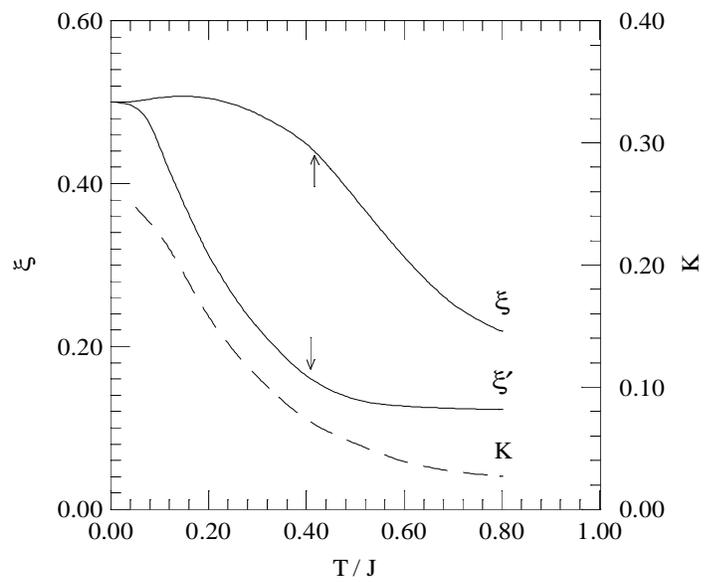
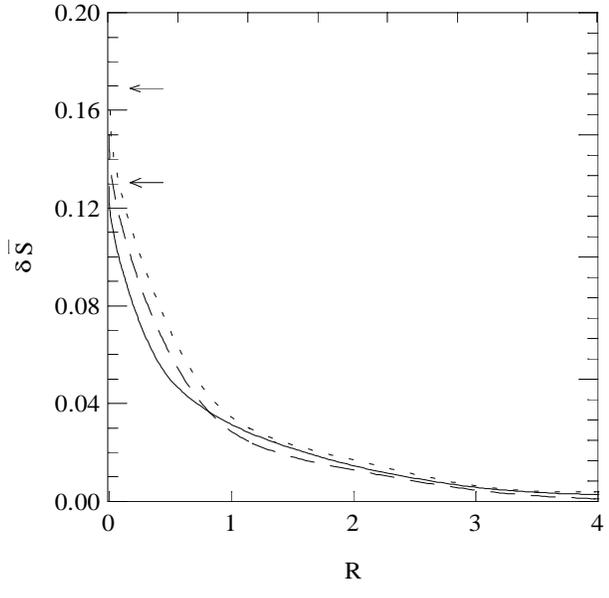
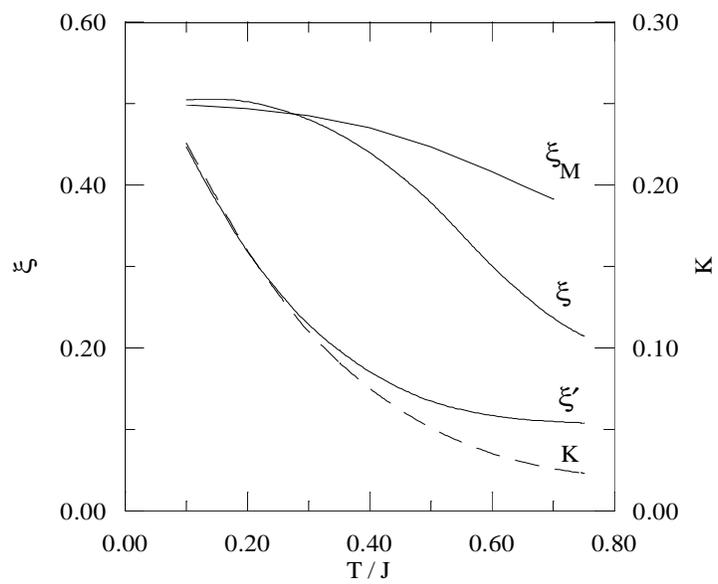
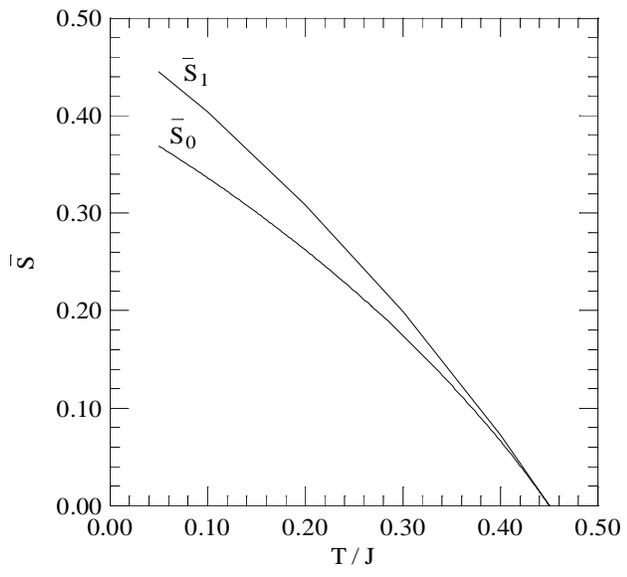
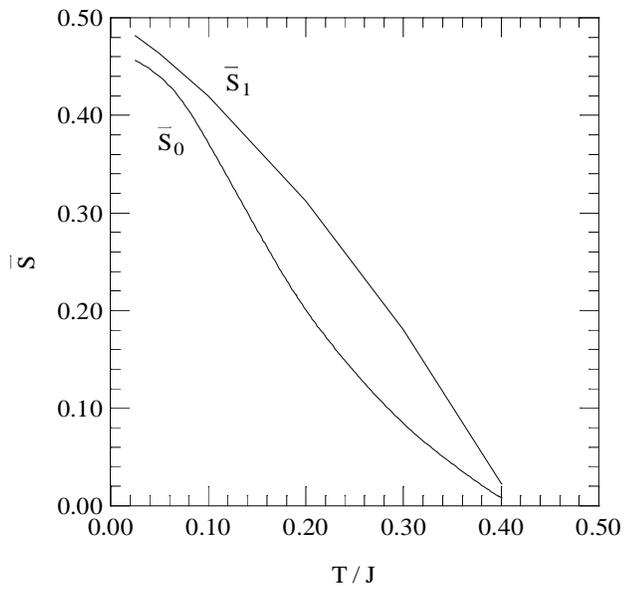
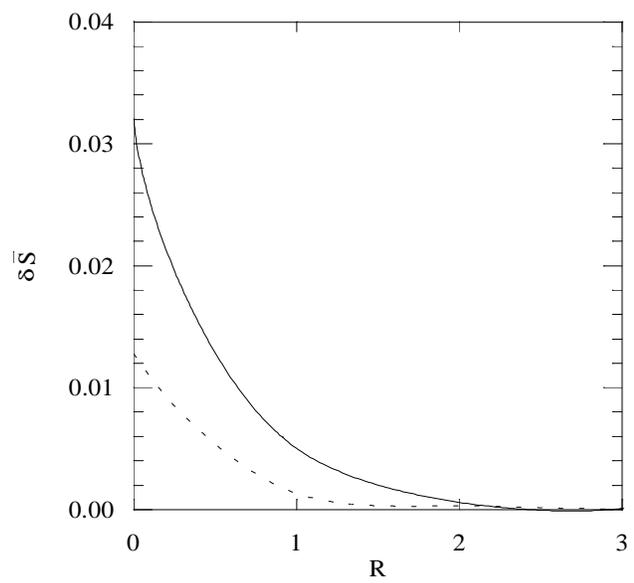
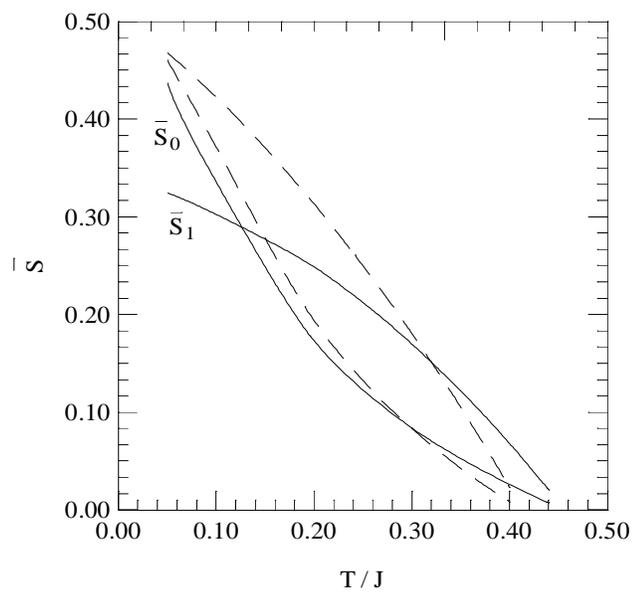
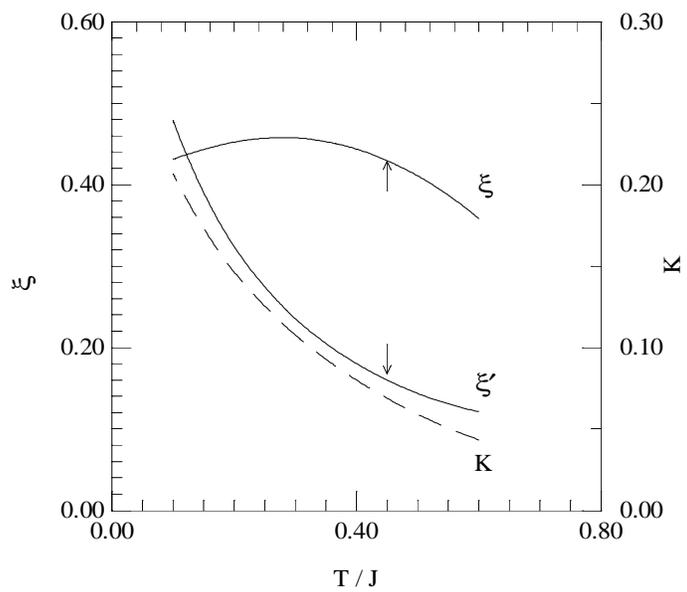
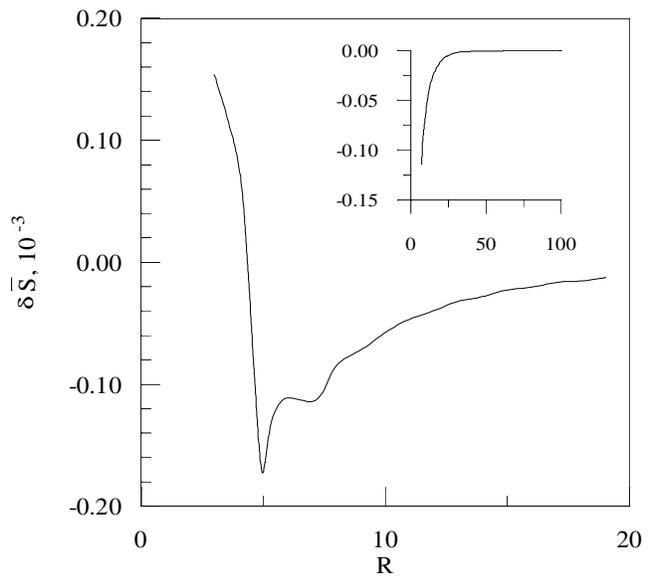
\end{document}